\documentclass[reprint,aps,prd,twocolumn,showpacs,nofootinbib,preprintnumbers,superscriptaddress]{revtex4-1}
\usepackage[T1]{fontenc}
\usepackage[latin9]{inputenc}
\usepackage{bm}
\usepackage{amsmath}
\usepackage{esint}
\usepackage{color}
\usepackage{graphicx}% Include figure files
\PassOptionsToPackage{normalem}{ulem}
\usepackage{ulem}

\usepackage{times}
\usepackage{hyperref}

\begin{document}

\title{Unifying framework for scalar-tensor theories of gravity}

\author{Xian Gao}%
    \email[Email: ]{gao@th.phys.titech.ac.jp}
    \affiliation{%
        Department of Physics, Tokyo Institute of Technology,\\ 
        2-12-1 Ookayama, Meguro, Tokyo 152-8551, Japan}

\date{June 4, 2014}

\begin{abstract}
A general framework for effective theories propagating two tensor and one scalar degrees of freedom is investigated.
Geometrically, it describes dynamical foliation of spacelike hypersurfaces coupled to a general background, in which the scalar mode encodes the fluctuation of the hypersurfaces.  
Within this framework, various models in the literature---including $k$-essence, Horndeski theory, the effective field theory of inflation, ghost condensate as well as the Ho\v{r}ava gravity---get unified.
Our framework generalizes the Horndeski theory in the sense that, it propagates the correct number of degrees of freedom, although the equations of motion are generally higher order. 
We also identify new operators beyond the Horndeski theory, which yield second order equations of motion for linear perturbations around an a Friedmann-Robertson-Walker background.
\end{abstract}

\maketitle

% % % % % % % % % % % % %
\section{Introduction}
Models of inflation and dark energy are mainly based on degrees of freedom beyond general relativity (GR). 
These additional degrees of freedom are most straightforwardly realized by scalar fields.
Over the years, $k$-essence \cite{ArmendarizPicon:1999rj} was studied as the most general local theory for scalar fields, which involves at most first derivatives of the fields in the Lagrangian.
Until recently, this understanding was promoted to higher order in derivatives, by rediscovering the Horndeski theory \cite{Horndeski:1974wa}---the most general covariant scalar-tensor theory involving up to second derivatives in the Lagrangian, while still leading to second order equations of motion for both scalar field and the metric---as the ``generalized Galileon'' \cite{Deffayet:2011gz}, which 
generalizes the flat-space ``Galileon'' \cite{Nicolis:2008in}  as well as its covariantization \cite{Deffayet:2009wt,Deffayet:2009mn} and includes the Dvali-Gabadadze-Porrati (DGP) model \cite{Dvali:2000hr} as a special case.
The ``second-order'' nature of Horndeski theory prevents it from extra ghostlike degrees of freedom and instabilities.

On the other hand, new degrees of freedom may arise when symmetries are reduced.
Therefore, an alternative approach to these additional degrees of freedom beyond GR is to construct theories which do not respect the full diffeomorphism of GR.
A well-studied example in this approach is the effective field theory (EFT) of inflation \cite{Creminelli:2006xe,Cheung:2007st} (which showed its first appearance in  ghost condensate \cite{ArkaniHamed:2003uy}), which describes the fluctuations around a time evolution background.
Another example, although initially motivated by a different purpose, is (more precisely, the nonprojectable version of) Ho\v{r}ava gravity \cite{Horava:2009uw}, where a preferred foliation structure of spacetime is introduced.
In both cases, the full spacetime symmetry is spontaneously broken to the reduced time-dependent spatial diffeomorphism on the hypersurfaces.

Now there are two different but equivalent formulations of theories.
One is in terms of scalar field(s) (``$\phi$-language''), the other is in terms of extrinsic and intrinsic geometric quantities associated with the hypersurfaces, which we refer to as ``brane-language'' for short.
The classification of various models is thus illustrated below:
	\begin{center}
	\begin{tabular}{ccc}
	``$\phi$-language'' & $\quad\Longleftrightarrow\quad$ & ``brane-language''\tabularnewline
	\cline{1-1} \cline{3-3} 
	$k$-essence &  & EFT of inflation\tabularnewline
	DGP &  & Ho\v{r}ava (non-proj.)\tabularnewline
	Horndeski &  & ghost condensate \tabularnewline
	\end{tabular}
	\par\end{center}
We use ``brane-language'' simply to emphasize that it does not necessarily imply the unitary gauge, since one can always formulate the theory covariantly, as we do in this work.

Formulations in two languages can be ``translated'' into each other.
For example, while appearing to be nonrelativistic as being written in the unitary gauge, both the EFT of inflation and Ho\v{r}ava gravity can be viewed as the ``gauge-fixed'' version of some covariant theories, where full diffeomorphism can be restored using the St\"{u}ckelberg trick.
For the EFT of inflation, this has been performed in \cite{Cheung:2007st} by introducing the Goldstone field $\pi$.
Similarly, in \cite{Germani:2009yt,Blas:2009yd} (see also \cite{Blas:2009qj}) the Ho\v{r}ava gravity was reformulated in a fully covariant manner, as describing spacelike hypersurfaces specified by a scalar field $\phi$ coupled to a general background.
Conversely, the Horndeski theory can also be recast in terms of extrinsic and intrinsic curvatures in the ``brane-language'' \cite{Gleyzes:2013ooa}.

The ``brane-language'' has special advantages since in which the dynamical degrees of freedom are made transparent.
In the study of EFT of inflation, much attention was paid to polynomials of (perturbations of) lapse function $\delta g^{00} \equiv 2\delta N/N^3$ and the extrinsic curvature $\delta K^i_j$ with time-dependent parameters.
For the Ho\v{r}ava gravity, besides the linear combination $K_{ij}K^{ij}- \lambda K^2$, attention was mainly focused on higher order polynomials built of spatial curvature and its derivatives such as ${}^{(3)}R^2$, ${}^{(3)}R_{ij}{}^{(3)}R^{ij}$, etc., with constant parameters.
In the healthy extension of Ho\v{r}ava gravity \cite{Blas:2009qj}, terms such as $(\partial_i N)^2$ were also introduced.
On the other hand, couplings between the extrinsic and intrinsic curvatures [such as $K {}^{(3)}R$ and $K_{ij}{}^{(3)}R^{ij}$] and terms cubic in the extrinsic curvature naturally arise when writing the Horndeski theory in the brane-language \cite{Gleyzes:2013ooa}.
Very recently in \cite{Gleyzes:2014dya}, more general operators were introduced, where parameters are generalized as functions of $(t,N)$.
In this work, we take one step further and investigate a generic framework in the brane-language, within which various existing models can be unified.

% % % % % % % % % % % % % % % % % % %
\section{Framework}
The key ingredient in our construction is the foliation of codimension-one spacelike hypersurfaces, which is encoded into a scalar field $\phi$ with a timelike gradient. 
The normal vector to the foliation is $n_{a}=-N\nabla_{a}\phi$, with lapse $N$ as the normalization coefficient.
The components of $\nabla_b n_a$ parallel and perpendicular to the hypersurface correspond to the extrinsic curvature $K_{ab} = h^c_a \nabla_c n_b\equiv D_b n_a$  and the acceleration $a_a = n^b \nabla_b n_a \equiv D_a \ln N$ respectively, where  $h_{ab} = g_{ab} + n_an_b$ is the induced metric on the hypersurfaces, $D_a$ is the intrinsic covariant derivative compatible with  $h_{ab}$.
For the sake of simplicity, in the following $R$ and $R_{ab}$ denote the intrinsic Ricci scalar and tensor on the hypersurfaces, while curvature terms of the spacetime are denoted as ${}^{(4)}R$ and ${}^{(4)}R_{ab}$ etc.

We restrict ourselves to the case where intrinsic derivatives only act on intrinsic curvature terms. That is, we omit terms such as $D_c K_{ab}$, $D_a a_b$, etc., although which may generally be allowed and interesting.
We consider a class of Lagrangians of the following form
	\begin{equation}
		\mathcal{L}= \sum_{n=1} \mathcal{K}_n + \mathcal{V}, \label{L_gen}
	\end{equation}
with 
	\begin{equation}
		\mathcal{K}_n = \mathcal{G}_{(n)}^{a_{1}b_{1},\cdots,a_{n}b_{n}}K_{a_{1}b_{1}}\cdots K_{a_{n}b_{n}}, \label{L_kin}
	\end{equation}
where $\mathcal{G}_{(n)}$'s and $\mathcal{V}$ are general functions of
	\begin{equation}
		\left(\phi,N,h_{ab}, R_{ab},a_a, D_{a}\right). \label{var_gen}
	\end{equation}
When writing (\ref{L_kin}), the symmetries of indices $(a_ib_i)$ of $\mathcal{G}_{(n)}$'s are understood.
We do not include $R_{abcd}$, which is not an independent quantity since the spatial hypersurfaces are 3-dimensional.
Following the same strategy of \cite{Horava:2009uw}, it is convenient to view  $\mathcal{K}_n$ as the ``kinetic'' terms, since $K_{ab}\equiv \frac{1}{2} \pounds_{\bm{n}} h_{ab}$ while the Lie derivative $\pounds_{\bm{n}}$ with respect to $n^a$ plays the role of a ``time derivative,'' and $\mathcal{V}$ as the ``potential'' terms.

Comments are in order. 
First, we do not include the shift vector $N_a$,
which itself is not a genuine geometric quantity of the foliation. 
Instead, it merely characterizes the gauge freedom of choosing a time direction  through $t_a = N n_a + N_a$. 
In fact, blindly including terms such as $N_a N^a$ would inevitably introduce unwanted degrees of freedom.
Second, coefficients of the kinetic terms $\mathcal{G}_{(n)}$'s and potential terms $\mathcal{V}$ have functional dependence on $N$, even nonlinearly. 
As we shall see, this will crucially ensure the health of our construction. 
Indeed, this is the idea of introducing nonlinear terms of $a_a$ in the healthy extension of Ho\v{r}ava gravity \cite{Blas:2009qj}.
Moreover, our construction is closely related to the Einstein-aether theory \cite{Jacobson:2000xp}, which is an effective theory describing a timelike unit vector field coupled to gravity.
The main difference is that, in our formalism, the unit vector $n_a$ is hypersurface orthogonal.

In the following we propose a ``cubic construction'' as an explicit example of our general setup (\ref{L_gen})-(\ref{L_kin}), by imposing two further restrictions: 
(i) there are no higher order derivatives in the Lagrangian when going into ``$\phi$-language,'' i.e., we omit terms such as $\Delta R$, $(D_c R_{ab})^2$ etc and 
(ii) the number of second order derivative operators does not exceed three.
This allows us to exhaust all the possible operators: 
for the ``kinetic terms''
	\begin{eqnarray}
	\mathcal{K}_{1} & = & \left(a_{0}+a_{1}R+a_{3}R^{2}+a_{4}R_{ab}R^{ab}+a_{5}a_{a}a^{a}\right)K\nonumber \\
	 &  & +\left[\left(a_{2}+a_{6}R\right)R^{ab}+a_{7}R_{c}^{a}R^{bc}+a_{8}a^{a}a^{b}\right]K_{ab},\label{L_min_1}\\
	\mathcal{K}_{2} & = & \left(b_{1}+b_{3}R\right)K^{2}+\left(b_{2}+b_{4}R\right)K_{ab}K^{ab}\nonumber \\
	 &  & +\left(b_{5}KK_{ab}+b_{6}K_{ac}K_{b}^{c}\right)R^{ab},\label{L_min_2}\\
	\mathcal{K}_{3} & = & c_{1}K^{3}+c_{2}KK_{ab}K^{ab}+c_{3}K_{b}^{a}K_{c}^{b}K_{a}^{c},\label{L_min_3}
	\end{eqnarray}
and for the ``potential terms''
	\begin{eqnarray}
	\mathcal{V} & = & d_{0}+d_{1}R+d_{2}R^{2}+d_{3}R_{ab}R^{ab}+d_{4}a_{a}a^{a}\nonumber \\
	 &  & +d_{5}R^{3}+d_{6}RR_{ab}R^{ab}+d_{7}R_{b}^{a}R_{c}^{b}R_{a}^{c}\nonumber \\
	 &  & +d_{8}Ra_{a}a^{a}+d_{9}R_{ab}a^{a}a^{b},\label{L_min_V}
	\end{eqnarray}
where $a_n,b_n,c_n,d_n$ are general functions of $(\phi,N)$.
As we shall see, this ``cubic construction'' has virtually included \emph{all} previous models, while still possessing new interesting extensions.
The ``6-parameter'' Lagrangian in \cite{Gleyzes:2014dya} corresponds to
	\begin{align*}
	 & a_{0}=A_{3},\quad-2a_{1}=a_{2}=B_{5},\quad b_{1}=-b_{2}=A_{4},\\
	 & c_{1}=-\frac{1}{3}c_{2}=\frac{1}{2}c_{3}=A_{5},\quad d_{0}=A_{2},\quad d_{1}=B_{4},
	\end{align*}
with all other coefficients vanishing.

\section{Hamiltonian analysis}
We wish to show our theory (\ref{L_gen})-(\ref{L_kin}) is healthy in the sense that it does not propagate unwanted degree(s) of freedom other than the two tensor and one scalar modes.
Counting number of degrees of freedom can be well performed in the Hamiltonian analysis.
To this end, we choose the unitary gauge with $t=\phi$, which corresponds to the coordinates adapted to the foliation structure, i.e., the well-known Arnowitt-Deser-Misner coordinates, where the conjugate momenta of the spatial metric $h_{ij}$ are given by
	\begin{equation}
	\frac{2}{\sqrt{h}}\pi^{ij}=\mathcal{G}_{(1)}^{ij}+\sum_{n=1}\left(n+1\right)\mathcal{G}_{(n+1)}^{ij,k_{1}l_{1},\cdots,k_{n}l_{n}}K_{k_{1}l_{1}}\cdots K_{k_{n}l_{n}}. \label{conj_mom}
	\end{equation}
Generally (\ref{conj_mom}) is a nonlinear algebraic equation for $K_{ij} \equiv  \frac{1}{2N}\big(\partial_t{h}_{ij} - 2\nabla_{(i} N_{j)}\big)$.
In case $K_{ij}$ cannot be fully solved in terms of $\pi_{ij}$, additional primary constraints are present, which further reduce the phase space and may cause pathological problems.
In our construction, $\mathcal{K}_{2}$ acts as a ``quadratic kinetic term,'' which we require to be not degenerate, i.e., we assume $\mathcal{G}_{(2)}^{ij,kl}$ possesses an inverse  $\mathcal{G}_{(2)ij,kl}^{-1}$ satisfying $\mathcal{G}_{(2)}^{ij,k'l'}\mathcal{G}_{(2)k'l',kl}^{-1}=\delta_{(k}^{i}\delta_{l)}^{j}$.
In this case, one may in principle solve $K_{ij}$ in terms of (e.g., a series of) $\pi^{ij}$, which can be well performed locally in some nonsingular branch of $K_{ij}$.
Fortunately, the explicit solution  is not needed for our purpose.
By definition and simple manipulations, the Hamiltonian takes the general form
	\begin{equation}
		\mathcal{H} \simeq N \mathcal{C} +N_{i}\mathcal{C}^{i}, \label{Ham}
	\end{equation}
where $\mathcal{C}^{i}=-2\sqrt{h}\nabla_{j}\left(h^{-1/2}\pi^{ij}\right)$ are exactly the same three momentum constraints generated by $N_i$ as in GR, and
	\begin{equation}
		\mathcal{C} = 2\pi^{ij}K_{ij}-\sqrt{h}\mathcal{L}. \label{C_cons}
	\end{equation}
In (\ref{Ham}) and (\ref{C_cons}), $K_{ij}$ should be thought of as a function of $\pi^{ij}$:
	\begin{equation}
		K_{ij}=K_{ij}\left(\pi^{kl},t,N,h_{kl},R_{kl},a_{k}\right), \label{K_pi}
	\end{equation}
which has nothing to do with the shift $N_i$.

Now comes the crucial point.
If all $\mathcal{G}_{(n)}$'s, $\mathcal{V}$ and thus the solution (\ref{K_pi}) have no functional dependence on the lapse $N$, so does $\mathcal{C}$ defined in (\ref{C_cons}). 
In this case $N$ enters the Hamiltonian linearly and acts as a Lagrange multiplier.
This is exactly the case of GR, where $N$ generates a first class constraint $\mathcal{C}=0$.
A subtle example, however, is the original version of Ho\v{r}ava gravity \cite{Horava:2009uw}, where although $N$ still acts as a Lagrange multiplier, the corresponding constraint is not first class any more and the dimension of the phase space is shown to be odd \cite{Li:2009bg}.
This pathological behavior was cured in \cite{Blas:2009qj} by adding invariants of acceleration such as $(a_i a^i)^n$ in the potential terms.
Since $a_i = \partial_i \ln N$, this is essentially to add nonlinear functional dependence on $N$ in the Hamiltonian (since $\partial \mathcal{C}/ \partial N \neq 0$), which prevents $N$ from being a Lagrange multiplier.
This is also the case for our general construction (\ref{L_gen})-(\ref{L_kin}).
As long as $\mathcal{G}_{(n)}$'s and/or $\mathcal{V}$ depend on $N$, $\mathcal{C}$ defined in (\ref{C_cons}) acquires functional dependence on $N$, and thus a new pair of degrees of freedom arises in the phase space, which corresponds to the scalar mode.

% % % % % % % % % % % % % % % % % %
\section{Dictionary}
As we have stated before, theories in the $\phi$-language and the brane-language can be explicitly translated into each other.
For example, the extrinsic curvature and the acceleration are written in the $\phi$-language as
	\begin{eqnarray}
	K_{ab} & = & \frac{1}{\left(2X\right)^{5/2}}\Big[-4X^{2}\nabla_{a}\nabla_{b}\phi\nonumber \\
	 &  & +\nabla_{a}\phi\nabla_{b}\phi\nabla^{c}\phi\nabla_{c}X+4X\nabla_{(a}\phi\nabla_{b)}X\Big],\label{Kab_phi}\\
	a_{a} & = & -\frac{1}{4X^{2}}\left(\nabla_{a}\phi\nabla^{b}\phi\nabla_{b}X+2X\nabla_{a}X\right),\label{aa_phi}
	\end{eqnarray}
respectively, where $X=-(\nabla\phi)^2/2$. We also have $N=1/\sqrt{2X}$ and $h_{ab}=g_{ab}+\frac{1}{2X}\nabla_{a}\phi\nabla_{b}\phi$.
Using these relations (as well as the Gauss-Codazzi-Ricci equations), the $\phi$-Lagrangian corresponding to (\ref{L_gen})-(\ref{L_kin})  can be easily derived, from which various terms beyond the Horndeski theory while still having healthy behaviors can be read \cite{Gao:toappear}.

For derivatives of the scalar field, we have $\nabla_a \phi = -n_a/N$,
	\begin{equation}
		\nabla_{a}\nabla_{b}\phi=\frac{1}{N}\left(-n_{a}n_{b}\rho+2n_{(a}a_{b)}-K_{ab}\right), \label{nabla2_phi_dec}
	\end{equation}
where $\rho \equiv \pounds_{\bm{n}}\ln N$, and 
	\begin{eqnarray}
	 &  & \nabla_{a}\nabla_{b}\nabla_{c}\phi\nonumber \\
	 & = & \frac{1}{N}\Big[n_{a}n_{b}n_{c}\left(-\rho^{2}+\pounds_{\bm{n}}\rho-2a^{d}a_{d}\right)\nonumber \\
	 &  & \quad-2n_{a}n_{(b}\big(\pounds_{\bm{n}}a_{c)}-2a_{c)}\rho-2K_{c)}^{d}a_{d}\big)\nonumber \\
	 &  & \quad+n_{a}\big(-2a_{b}a_{c}-\rho K_{bc}+\pounds_{\bm{n}}K_{bc}-2K_{bd}K_{c}^{d}\big)\nonumber \\
	 &  & \quad-\left(\pounds_{\bm{n}}a_{a}-2a_{a}\rho-2K_{a}^{d}a_{d}\right)n_{b}n_{c}\nonumber \\
	 &  & \quad+2\left(-a_{a}a_{(b}+D_{a}a_{(b}-\rho K_{a(b}-K_{a}^{d}K_{d(b}\right)n_{c)}\nonumber \\
	 &  & \quad+a_{a}K_{bc}+2K_{a(b}a_{c)}-D_{a}K_{bc}\Big],\label{nabla3_phi_dec}
	\end{eqnarray}
while in this work we omit the decomposition of $\nabla_a \nabla_b \nabla_c \nabla_d \phi$ due to its length (see however \cite{Gao:toappear}).
By employing these relations, any $\phi$-Lagrangian can be written in the brane-language, as being performed to the Horndeski theory in \cite{Gleyzes:2013ooa}.
We emphasize that the covariant nature of our formalism enables us to write down terms in the brane-language in a covariant form.

% % % % % % % % % % % % % % % % % % %
\section{Linear perturbations}
When expanding around a spatially-flat Friedmann-Robertson-Walker (FRW) background, the perturbation theory becomes dramatically simple in our formalism, due to its ``spacetime-splitting'' nature.
Note $R_{ab}$ and $a_a$ start from linear order in perturbations, thus when considering linear perturbations, only a  limited number of operators in (\ref{L_min_1})-(\ref{L_min_V}) contribute.

There are two choices of coordinates, i.e., gauges.
One is popularly used in the study of EFT of inflation, where the scalar mode is pushed to the Goldstone field $\pi$ through the St\"{u}ckelberg trick $\frac{1}{N^2}\rightarrow g^{\mu\nu}\partial_{\mu}\left(t+\pi \right) \partial_{\nu}\left(t+\pi \right)$. 
This is essentially to work in the $\phi$-Language and perturb directly the $\phi$ field.
The other one is the unitary gauge, which we will employ below.
The perturbations of $N$ and $K_{ij}$ are parametrized by $N=e^{\alpha}$ and $K_{ij} = \frac{1}{2N}\big(\dot{h}_{ij} - 2\nabla_{(i} N_{j)}\big)$ with  $h_{ij} = a^2 e^{2\zeta} \left(e^{\gamma}\right)_{ij}$, where $a$ is the scale factor, $\left(e^{\gamma}\right)_{ij} \equiv \delta_{ij} + \gamma_{ij} + \frac{1}{2} \gamma_{ik}\gamma_{kj}+\cdots$ with $\gamma_{ij}$ the transverse and traceless tensor perturbation satisfying $\partial_i \gamma_{ij} = \gamma_{ii} = 0$ (repeated lower spatial indices are summed by $\delta^{ij}$).
Moreover, all the coefficients in (\ref{L_min_1})-(\ref{L_min_V}) are now functions of $t$ and $N$, e.g., $a_0 = a_0(t,N)$, etc.

We investigate a ``minimal'' version of (\ref{L_min_1})-(\ref{L_min_V}), with  
	\begin{equation}
		a_0, a_1, a_2, \quad b_1, b_2, \quad c_1, c_2, c_3, \quad d_0, d_1, \label{coeff_min}
	\end{equation} 
as arbitrary functions of $t$ and $N$, while all other coefficients are vanishing. This 10-parameter ``minimal'' version has already included the Horndeski theory \cite{Horndeski:1974wa,Deffayet:2011gz} and the  extension in \cite{Gleyzes:2014dya} as special cases. 
The background equation of motion is given by $\bar{\mathcal{E}}=0$ with
	\begin{equation}
		\bar{\mathcal{E}}=d_{0}+d_{0}'+3a_{0}'H-3(\lambda_{1}-\lambda_{1}')H^{2}-3(2\lambda_{2}-\lambda_{2}')H^{3}, \label{bg_eq}
	\end{equation}
with $\lambda_{1}\equiv3b_{1}+b_{2}$ and $\lambda_{2}\equiv9c_{1}+3c_{2}+c_{3}$,
where $H$ is the Hubble parameter, a prime ``$'$'' denotes derivative with respect to $N$, e.g., $a_{0}'\equiv\left.\frac{\partial a_{0}\left(t,N\right)}{\partial N}\right|_{N=1}$, etc.

The quadratic Lagrangian for the tensor perturbations reads (in momentum space)
	\begin{equation}
		\mathcal{L}_{2}^{\mathrm{T}}=\frac{a^{3}}{4}\left(\mathcal{G}_{\mathrm{T}}\dot{\gamma}_{ij}^{2}+\mathcal{W}_{\mathrm{T}}\frac{k^{2}}{a^{2}}\gamma_{ij}^{2}\right), \label{L2_T}
	\end{equation}
where
	\begin{eqnarray}
	\mathcal{G}_{\mathrm{T}} & = & b_{2}+3\left(c_{2}+c_{3}\right)H,\label{G_T}\\
	-\mathcal{W}_{\mathrm{T}} & = & d_{1}+\frac{3}{2}\left(2a_{1}+a_{2}\right)H+\frac{1}{2}\frac{\mathrm{d}a_{2}}{\mathrm{d}t},\label{W_T}
	\end{eqnarray}
with $H$ the Hubble parameter.
It is interesting that only 6 operators proportional to $a_1$, $a_2$, $b_2$, $c_2$, $c_3$ and $d_1$ contribute to the linear tensor perturbations.
Remarkably, with arbitrary combination of these 6 operators, which is definitely beyond the Horndeski theory and also the Lagrangian in \cite{Gleyzes:2014dya}, the equation of motion for the linear tensor perturbations stays at the second order, with propagating speed given by $c_{\mathrm{T}}^2 = \mathcal{W}_T/\mathcal{G}_T$.
We require $\mathcal{G}_\mathrm{T}>0$ to avoid the ghost instability, while  $\mathcal{W}_\mathrm{T}$ may be negative during some time interval and cross zeros at some points, as long as the instabilities do not become too large and invalidate the perturbation theory.

The quadratic Lagrangian for the scalar perturbation $\zeta$ is
	\begin{equation}
		\mathcal{L}_{2}^{\mathrm{S}}=a^{3}\left[\mathcal{G}_{\mathrm{S}}\dot{\zeta}^{2}+\left(\mathcal{W}_{\mathrm{S}}^{(0)}+\mathcal{W}_{\mathrm{S}}^{(1)}\frac{k^{2}}{a^{2}}\right)\frac{k^{2}}{a^{2}}\zeta^{2}\right], \label{L2_S}
	\end{equation}
with
	\begin{eqnarray}
	\mathcal{G}_{\mathrm{S}} & = & \frac{1}{3\Xi}\left[6\Gamma_{1}^{2}\Gamma_{3}-\left(2\Gamma_{1}-3C_{2}\right)\bar{\mathcal{E}}_{,H}^{2}\right]+3\Gamma_{1},\label{GS}\\
	\mathcal{W}_{\mathrm{S}}^{(0)} & = & 2\Gamma_{2}+\frac{1}{a}\frac{\mathrm{d}}{\mathrm{d}t}\Big\{\frac{a}{9\Xi}\big[12\left(\Gamma_{1}-3C_{2}\right)\bar{\mathcal{E}}_{,H}\left(d_{1}+\Gamma_{2}'\right)\nonumber \\
	 &  & \qquad+3C_{1}\left(\bar{\mathcal{E}}_{,H}^{2}-6\Gamma_{1}\Gamma_{3}\right)\big]-2a\Gamma_{2,H}\Big\},\label{WS0}\\
	\mathcal{W}_{\mathrm{S}}^{(1)} & = & \frac{2}{3\Xi}\big[C_{1}\left(3\Gamma_{3}C_{1}-4\bar{\mathcal{E}}_{,H}\left(d_{1}+\Gamma_{2}'\right)\right)\nonumber \\
	 &  & \qquad+24C_{2}\left(d_{1}+\Gamma_{2}'\right)^{2}\big],\label{WS1}
	\end{eqnarray}
where
$\Xi\equiv\frac{1}{9}\bar{\mathcal{E}}_{,H}^{2}-2\Gamma_{3}C_{2}$,
$\Gamma_1 \equiv \lambda_{1}+3\lambda_{2}H$, 
$\Gamma_{2}=d_{1}+\left(3a_{1}+a_{2}\right)H$, 
$\Gamma_{3}\equiv d_{0}+3d_{0}'+d_{0}''+3\left(a_{0}'+a_{0}''\right)H+3\left(\lambda_{1}-\lambda_{1}'+\lambda_{1}''\right)H^{2}+3\left(4\lambda_{2}-3\lambda_{2}'+\lambda_{2}''\right)H^{3}$,
$\bar{\mathcal{E}}_{,H}\equiv\partial\bar{\mathcal{E}}/\partial H$,
$\Gamma_{2,H}\equiv\partial\Gamma_2/\partial H$ 
and 
	\begin{equation}
		C_{1}\equiv2a_{1}+a_{2},\quad C_{2}=b_{1}+b_{2}+\left(9c_{1}+5c_{2}+3c_{3}\right)H.
	\end{equation}
We emphasize that, in deriving (\ref{L2_S}) coefficients in (\ref{coeff_min}) are assumed to depend on $N$ generally. Instead, as in the (nonprojectable) Ho\v{r}ava gravity, $\zeta$ loses its quadratic kinetic term and becomes nondynamical at linear order around a FRW background. However, time derivatives of $\zeta$ reappears linearly at higher orders, which not only implies the odd dimensionality of the phase space but also the strong coupling problem \cite{Blas:2009yd}. This pathological behavior is avoided in our general construction, as long as the coefficients in (\ref{coeff_min}) are generally functions of $N$. This fact is also consistent with the argument on the constraint analysis. 

Due to the presence of $\mathcal{W}_{\mathrm{S}}^{(1)}$, generally, the scalar mode acquires a nonrelativistic dispersion relation as in the ghost condensate \cite{ArkaniHamed:2003uy}.
From (\ref{L2_S}), the absence of ghost instability requires $\mathcal{G}_{\mathrm{S}}>0$ while $\mathcal{W}_{\mathrm{S}}^{(0)}+\mathcal{W}_{\mathrm{S}}^{(1)}k^{2}/a^{2}$ may be  negative during some time period.
Requiring $\mathcal{W}_{\mathrm{S}}^{(1)}=0$ and thus $C_1=C_2=0$ yields 3 constraints among 10 parameters:
	\begin{equation}
		2a_{1}+a_{2}=b_{1}+b_{2}=9c_{1}+5c_{2}+3c_{3}=0. \label{cons_s}
	\end{equation}
For Horndeski theory  and the extension in \cite{Gleyzes:2014dya}, all three constraints are satisfied and thus there are no higher spatial derivatives.
The first constraint in (\ref{cons_s}) fixes terms linear in $K_{ab}$ to be $G^{ab}K_{ab}$ with $G^{ab}$ the Einstein tensor, 
the second one fixes terms quadratic in $K_{ab}$ to be the Galileon-type $\sim \left(K^{2}-K_{ab}K^{ab} \right)$.
However, the last constraint in (\ref{cons_s}) implies for terms cubic in $K_{ab}$, besides the Galileon-type combination $\sim \left(K^{3}-3KK_{ab}K^{ab}+2K_{b}^{a}K_{c}^{b}K_{a}^{c}\right)$, there is another combination
	\begin{equation}
		\sim c(\phi,N)\left(3KK_{ab}K^{ab}-5K_{b}^{a}K_{c}^{b}K_{a}^{c}\right), \label{K3_new}
	\end{equation}
which also yields second order equations of motion for linear perturbations.
The existence of (\ref{K3_new}) is because higher order polynomials in $K_{ab}$ are degenerate for linear perturbations. In fact, expanding $\mathcal{K}_3$ at the quadratic order in $\delta K_{ij}\equiv K_{ij} - H\delta_{ij}$ yields $\frac{H}{a^{6}}\big[\left(9c_{1}+2c_{2}\right)\left(\delta K\right)^{2}+3\left(c_{2}+c_{3}\right)\delta K_{ij}\delta K_{ij}\big]$, while according to the EFT of inflation \cite{Cheung:2007st}, the cancellation of higher spatial derivative requires $9c_{1}+2c_{2}=-3\left(c_{2}+c_{3}\right)$, which is just the last constraint in (\ref{cons_s}).
To summarize, within the ``minimal version'' with coefficients (\ref{coeff_min}), we arrive at a ``7-parameter'' family of Lagrangians beyond the ``6-parameter'' one in \cite{Gleyzes:2014dya}, which yields second order equations of motion for linear perturbations. Of course, higher spatial derivatives will reappear on nonlinear orders, since the unique theory which has second order equations of motion to all orders is the Horndeski theory.
When going beyond this minimal version and switching on operators such as $a_a a^a$, higher order spatial derivatives will also  appear for both tensor and scalar modes \cite{Gao:toappear}.
Finally, (\ref{L2_S}) implies $\zeta$ is conserved on large scales when $\mathcal{W}_{\mathrm{S}}^{(0)}+\mathcal{W}_{\mathrm{S}}^{(1)}k^{2}/a^{2}\neq 0$. Following the same approach in \cite{Gao:2011mz}, one can show that $\zeta$ is conserved at fully nonlinear orders, even for the general construction (\ref{L_gen})-(\ref{L_kin}).

% % % % % % % % % % % %
\section{Conclusion}
We investigated a general framework for scalar-tensor theories (\ref{L_gen})-(\ref{L_kin}), which can be viewed as dynamical spacelike hypersurfaces coupled to a general background.
Different models in the literature, including the Horndeski theory, EFT of inflation, Ho\v{r}ava gravity, etc., now get unified as special cases of our general formalism.

Our framework generalizes the Horndeski theory by introducing higher order derivatives in a special manner. 
There exists a particular choice of coordinates adapted to the foliation, where higher order spatial derivatives are allowed while the temporal derivatives are kept up to the second order in the equations of motion, thus
the Cauchy problem with the correct number of initial data is manifest.
In a general frame, higher order spatial derivatives are transferred into higher order time derivatives, thus apparently additional degrees of freedom arise, which however, can be shown to be unphysical \cite{Blas:2009yd}. 
This is also reminiscent of the ghost free massive gravity \cite{deRham:2010kj}, where helicity modes apparently possess higher order equations of motion when going beyond the decoupling limit.
As a by-product of our general construction, similar to the investigation in \cite{Gleyzes:2013ooa,Gleyzes:2014dya}, we identify a new combination (\ref{K3_new}) which does not belong to the Horndeski theory, but still yields second order equations of motion for linear perturbations.
Our formalism also generalizes the EFT of inflation approach. 
Especially, we directly work with fully nonlinear operators instead of treating background/perturbation separately, which enables us to investigate nonperturbative solutions such as black holes.

% % % % % % % % % % % % % %
% % % % % % % % % % % %
\acknowledgments
I would like to thank C. Lin, S. Mukohyama and M. Yamaguchi for useful discussions.
I was supported by JSPS Grant-in-Aid for Scientific Research No. 25287054.

% % % % % % % % % % % % % %

\bibliography{Gao}

\end{document}